# Plasmonic-photonic crystal coupled nanolaser

**Taiping Zhang[1], Ségolène Callard[1], Cécile Jamois[1], Céline Chevalier[1], Di Feng[1,2], and Ali Belarouci[1]**

[1] Institut des Nanotechnologies de Lyon (INL), UMR 5270 CNRS-ECL-INSA-UCBL, Université de Lyon, Ecole Centrale de Lyon, 36 Avenue Guy de Collongue, Ecully, 69134, France

[2] School of Instrumentation Science and Optoelectronics Engineering, Beihang University, Beijing 100191, China

E-mail : Ali.Belarouci@ec-lyon.fr

**Abstract.** We propose and demonstrate a hybrid photonic-plasmonic nanolaser that combines the light harvesting features of a dielectric photonic crystal cavity with the extraordinary confining properties of an optical nano-antenna. In that purpose, we developed a novel fabrication method based on multi-step electron-beam lithography. We show that it enables the robust and reproducible production of hybrid structures, using fully top down approach to accurately position the antenna. Coherent coupling of the photonic and plasmonic modes is highlighted and opens up a broad range of new hybrid nanophotonic devices.

## 1. Introduction

Over the past 10 years we have witnessed a burst in the number and diversity of nanophotonics applications [1-5]. The downscaling of photonic devices that can efficiently concentrate the optical field into a nanometer-sized volume holds great promise for many emerging applications requiring tight confinement of optical fields, including information and communication technologies [6], sensors [7], enhanced solar cells [8] and lighting [9]. Photonic crystals [10, 11] are key geometries that have been studied for many years in this respect, as they provide full control over the dispersion of light by engineering the momentum of light $k(\omega)$. Furthermore, photonic crystal cavities [12] have shown strong confinement of light, leading to nanoscale on-chip lasers [13] light emitting diodes [14] and platforms to probe the quantum interaction between light and matter [15]. In a photonic crystal the sub-wavelength scale modulation of the dielectric constant results in large spatial and frequency variations of the LDOS, corresponding to either strongly inhibited or enhanced light emission at well-defined emitter locations and frequencies [16]. Although conventional dielectric cavities can be designed to have ultrahigh Q factors, the physical size of the optical mode is diffraction-limited and is always larger than the single cubic wavelength $(\lambda/n)^3$. In photonics and optoelectronics, metals were for decades perceived as being rather dull, devoid of interesting or useful optical properties that could be harnessed for optical components and devices. They have recently attracted considerable interest in the framework of plasmonics [17]. Ultrasmall mode volumes and high local field enhancement are achieved by exploiting the surface plasmon resonances of metallic nanostructures [18]. The capability of plasmonic systems to control light-matter interaction at the sub-wavelength scale offers remarkable opportunities to generate many novel concepts and applications in nanophotonics [19, 20]. However practical implementations have been hampered by a limited resonance strength (or confinement time) as a result of optical losses induced by metal absorption and radiation to free space continuum. To overcome this challenge we propose a new hybrid photonic-plasmonic nanocavity that exploits the coupling between localized surface plasmon resonance (LSPR) of a bowtie antenna and a photonic mode provided by an active photonic crystal structure. The working principle combines the light harvesting ability of the high-Q dielectric microcavity with the extraordinary confining properties of the plasmonic element. Although integrating the two systems on a single hybrid platform is at present challenging, it constitutes a powerful route to produce novel functional devices with synergetic assets.

Recently, hybrid systems where a plasmonic element is coupled to a photonic structure have been proposed by several groups [21-25]. Among the directions investigated to fabricate such hybrid structures, top-down technology in combination with nano-assembly has been actively explored. The two major issues are simultaneously controlling the relative position and orientation of the two constituents. These turn out to be particularly crucial when dealing with complex structures involving more than one plasmonic building block. Some of the approaches that have been successfully implemented to position metallic nanoparticles on photonic devices are exploiting nanomanipulation [22] based on atomic force microscopy (AFM) or optical trapping by the photonic element [25]. However, these methods are time consuming and offer only partial control over the location and the orientation of the optical antenna. In particular, the nano-antenna orientation is extremely important owing to its high polarization sensitivity. A method that tackles these challenges would be highly attractive. In the present paper, we present our approach to efficiently fabricate and couple a single plasmonic nano-antenna to a photonic crystal (PC) nanocavitiy. We show that the resulting hybrid photonic-plasmonic structure yields a mode allying the best of each world: the high quality factor provided by the PC cavity, together with the strong spatial confinement of the plasmonic nano-antenna. The fabrication process is based on lithography alignment, enabling for a well-controlled and reproducible nano-antenna positioning. Although the potential of this method is demonstrated at laboratory scale using electron-beam lithography, this fully top-down approach could be easily adapted to other lithography techniques suitable for larger scale, higher throughput and CMOS-compatible integration of hybrid devices.

**2. Structure design**

The basic building blocks of the hybrid cavity structure are sketched in Fig. 1. It consists of two elements: a high-Q PC microcavity, that has been described and studied in former works [26] and a bowtie nano-antenna. The PC microcavity is designed to present a high-Q mode (5800) at telecom wavelength (Fig. 1(c)). The optical response of the structure was modelled using 3D finite-difference time domain method (3D-FDTD) with perfectly matched layers boundary conditions. The nanocavity is excited with an oscillating dipole (pulsed temporal regime) properly oriented and positioned within the structure. Computational meshes were 10 nm for x, y, and z directions. The calculated intensity distributions of the $E_x$ and $E_y$ components of the field are shown in Fig. 1(b). In particular, they show that the maxima of $E_x$ and $E_y$ do not overlap spatially: $E_x$ maxima are present on the side of the cavity and $E_y$ maxima in the center of the cavity. This point is crucial for the design of the hybrid structure [27], as will be discussed later. The optical bowtie antenna (NA) consists of two coupled gold triangles separated by a nanometer size gap (20 nm) (Fig. 1(d)). Geometrical parameters have been optimized to provide good wavelength matching between the resonant mode of the optical NA and the PC. As highlighted in the FDTD simulations shown in Fig. 1(e), the NA strongly confines the electromagnetic field in the gap as soon as its polarization is aligned along the gap: in the following this direction will be referred as NA axis (Oy in Fig. 1). The other field component polarized perpendicularly to the NA axis is distributed at the four external corners of the bowtie triangles. Conversely to the PC cavity mode spectrum, the NA resonant mode spectrum is broad, allowing for an easy spectral overlap between the modes of each resonator.

In order to optimize the optical coupling between the fundamental mode of the microcavity and the NA resonance mode, the two modes must coincide spectrally as well as spatially (in position and polarization). Theoretical calculations helped to determine the relevant possible locations for the NA. Three of them have been selected and are presented in Fig. 2. Several positions are possible, as illustrated in Fig. 2: Fig. 2(a) and 2(b) show two optimal cases of hybrid structures, for which the polarization of the field is parallel to the y direction (with the NA in the centre of the cavity), or parallel to the x direction (with the NA at the edge of the cavity), respectively. In both cases, the NA is positioned exactly at a field maximum of the photonic mode. Fig. 2(c) shows an example for which the polarizations of the NA and of the optical mode are orthogonal. In Fig. 2(a) and 2(b), the field of the resulting mode is highly confined within the NA gap, indicating efficient coupling between the NA and the photonic cavity and strong spatial redistribution of the mode. Conversely, in Fig. 2(c), the field intensity can be seen at the edges of the NA and spatial confinement is less efficient. This result demonstrates that accurate positioning and orientation of the NA within the photonic cavity is absolutely crucial to obtain the desired coupling conditions between the two structures.

**3. Fabrication process**

The photonic crystal structure is realized in an InP-based substrate, which consists of a thin InP slab including four InAsP quantum wells (QWs) that has been grown by solid-source molecular beam epitaxy (MBE); the thickness of the slab (250 nm) has been chosen to allow for single-mode operation, and the slab has been bonded on a silica substrate [28] to yield a better heat sink, a robust mechanical stability, and enable front-side and back-side optical pumping of the QWs. The photoluminescence occurs over a broad spectral range between 1250 nm and 1650 nm.

The fabrication process steps that have been optimized for the realization of the hybrid plasmonic-photonic crystal device are summarized in Fig. 3. The challenges in the fabrication include highly accurate electron-beam (e-beam) lithography, high-quality dry etching techniques and ultimate alignment procedures between photonic and plasmonic devices. The fabrication process can be divided into the following 3 steps: defining alignment marks (Fig. 3(b)), realizing the photonic crystal laser (Fig. 3(c)), and positioning the optical nano-antenna on the backbone of the photonic structure (Fig. 3(d)). The alignment is achieved via the use of metal marks, defined in the first step of the fabrication, on which the next layers of lithography are aligned. After fabrication of the robust gold marks on the substrate, the second step is the realization of the photonic crystal devices, defined via e-beam lithography and alignment in a positive resist (polymethylmethacrylate – PMMA A4) and transferred first into a silica hard mask and then into the InP membrane by two successive reactive ion etching processes based on $CHF_3$ and $CH_4/H_2$ mixture, respectively. The hard mask is subsequently removed by RIE. Finally, in a third step the individual metallic nano-antennas are deterministically positioned on the PC cavity by e-beam lithography alignment and writing in a resist double-layer (MMA and PMMA C2), followed by electron-beam evaporation of 40 nm gold with 4 nm Ti adherence layer, and a lift-off process. With this 3-steps process, bowtie nano-antennas with reproducible gap sizes down to 10 nm have been successfully fabricated, with accuracy on their position better than 10 nm, as illustrated in Fig. 4.

**4. Optical characterization**

To validate our fabrication protocol, we investigated the optical properties of basic linear micro-cavities and hybrid structures fabricated in the same conditions. Optical characterization is performed by micro-photoluminescence spectroscopy at room temperature with the optical set-up showed in Fig. 5(a). Samples are optically pumped, using a pulsed laser diode (LD, Wavelength = 808 nm). The pump beam is focused onto the structures under normal incidence with an achromatic objective lens (20X). The duty cycle of the pulse LD is about 10% (repetition rate 233 ns, pulse width 22.3 ns). The light emitted in free space by the structure is collected through the same objective lens. Emission spectra are recorded using a spectrometer (0.15 nm resolution) and an InGaAs photodetector array after passing an 1100 nm-long pass filter [29].

Normalized emission spectra of the hybrid structures are shown in Fig. 5(b). They are presented together with the spectrum of the PC cavity without NA. It confirms that the presence of the NA preserves the lasing effect, which means that the Q-factor of the hybrid mode is still high enough to maintain stimulated emission. For the same group of nanodevices, the variation of the laser peak intensity with effective incident pump power was also measured to evaluate the impact of the NA on the laser threshold (Fig. 5(c)). They show that the presence of the NA increases the laser threshold. For the hybrid structures, the laser thresholds are 130 µW and 100 µW for positions 1 and 2 respectively, compared to 45 µW for the structure without NA. Similar results have been observed for several hybrid structures, demonstrating that the laser thresholds of hybrid structures are higher than the bare PC cavity without NA. It confirms that the NA induces optical losses that impact the lasing threshold. The losses may be induced by metallic absorption and scattering [22]. Wavelength-shifts have also been observed for structures with NA with respect to the structure without NA. On Fig. 5(b), red-shifts are respectively 3.9 nm and 5 nm for devices with NA at $E_y$ position (1590.8 nm) and with NA at $E_x$ position (1591.9 nm), compared to the bare cavity (1586.9 nm). These shifts are too high to be inferred to the thermal dependence of the lasing wavelength with the pumping power. Indeed, a red-shift due to thermal effects occurs and depends linearly on the pumping power. It was estimated to 0.00075 nm/µW for the three structures and is too weak to explain the shifts observed in Fig. 5(b). This shift is therefore considered as coming from the presence of the NA. Our hypothesis is that the shift is induced by the modes coupling (NA mode and PC mode): in this case, it should depend on the relative wavelength position of each mode. Experiments performed on 5 other groups of structures with different hole radius showed that red-shifts or blue-shifts can be obtained. However, while the wavelength of the PC cavity mode is experimentally accessible with the empty cavity, the exact NA resonance wavelength cannot yet be measured with our set up and works are still in progress to validate our hypothesis.

These results strongly indicate an optical coupling between the PC cavity and the NA, resulting in a hybrid mode with a sufficient quality factor to preserve lasing. This is in agreement with our numerical prediction. A complete demonstration of the hybrid nature of the mode requires optical near-field investigation to highlight the spatial confinement of the lasing mode within the NA gap. This will be the object of a later publication.

In summary, we realized a new kind of active plasmonic-photonic hybrid nanodevices with a novel method based on multi-step lithography alignment. Compared to other related works [21], this method can be directly adapted to the fabrication of hybrid systems on a large scale, and enables the realization of hybrid structures with complex geometry as well. We showed that the hybrid mode preserves the spectral feature of the photonic PC mode and allows for laser emission at room temperature. Simulations predict a strong reduction of the mode volume induced by the presence of the plasmonic building block. The experimental demonstration of this feature by scanning optical near-field microscopy is underway. We believe that this hybrid platform may open the route to applications in integrated opto-plasmonic devices for quantum information processing, as efficient single photon sources or nanolasers, or as sensing element for biochemical species.

**Acknowledgments**

This work was performed at Nanolyon platform and was supported by CNRS and China Scholarship Council. Authors thank the technical staff for support and fruitful discussions. T. Zhang would like to acknowledge Dr. Feng Lin (Lawrence Berkeley National Laboratory) for his precious comments.

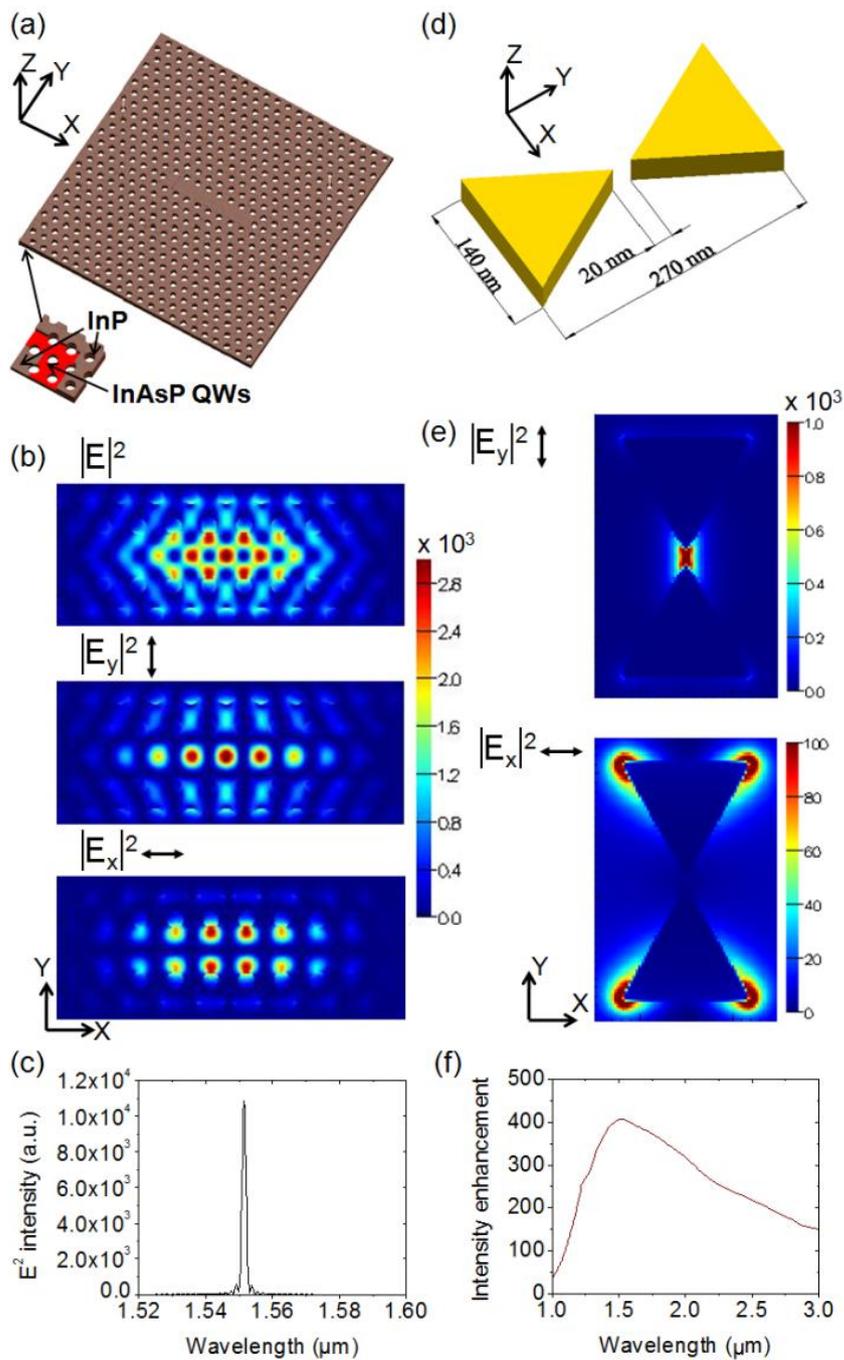

**Figure 1:** Basic building blocks illustrations of the hybrid cavity. (a) Active photonic crystal-based linear microcavity : 4 InAsP QWs are located in the InP membrane (b) Calculated electric field intensity distribution $|E|^2$, $|E_x|^2$, and $|E_y|^2$ inside the nanocavity at the resonant wavelength of the fundamental mode ($\lambda$=1.55 μm) (c) Spectrum of the nanocavity mode (d) Artistic view of a gold Bow-tie nanoantenna(NA) (e) Calculated electric field intensity distribution $|E_y|^2$ and $|E_x|^2$ of the resonant mode of the NA, (f) Spectrum of the NA.

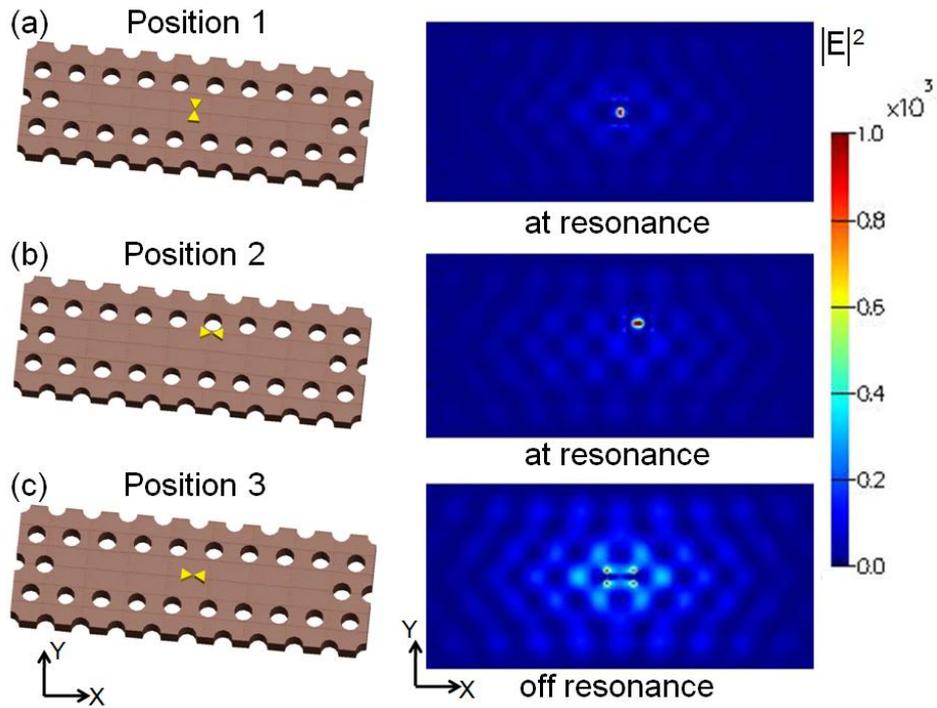

**Figure 2:** Three designs of hybrid nanostructures and their corresponding optical field distributions (FDTD simulations): (a) nano-antenna (NA) in the centre of the cavity and the polarization of the NA and the optical modes are parallel; (b) NA at the edge of the cavity and the polarization of the NA and the optical modes are parallel; (c) NA in the centre of the cavity and the polarization of the NA and the optical modes are orthogonal.

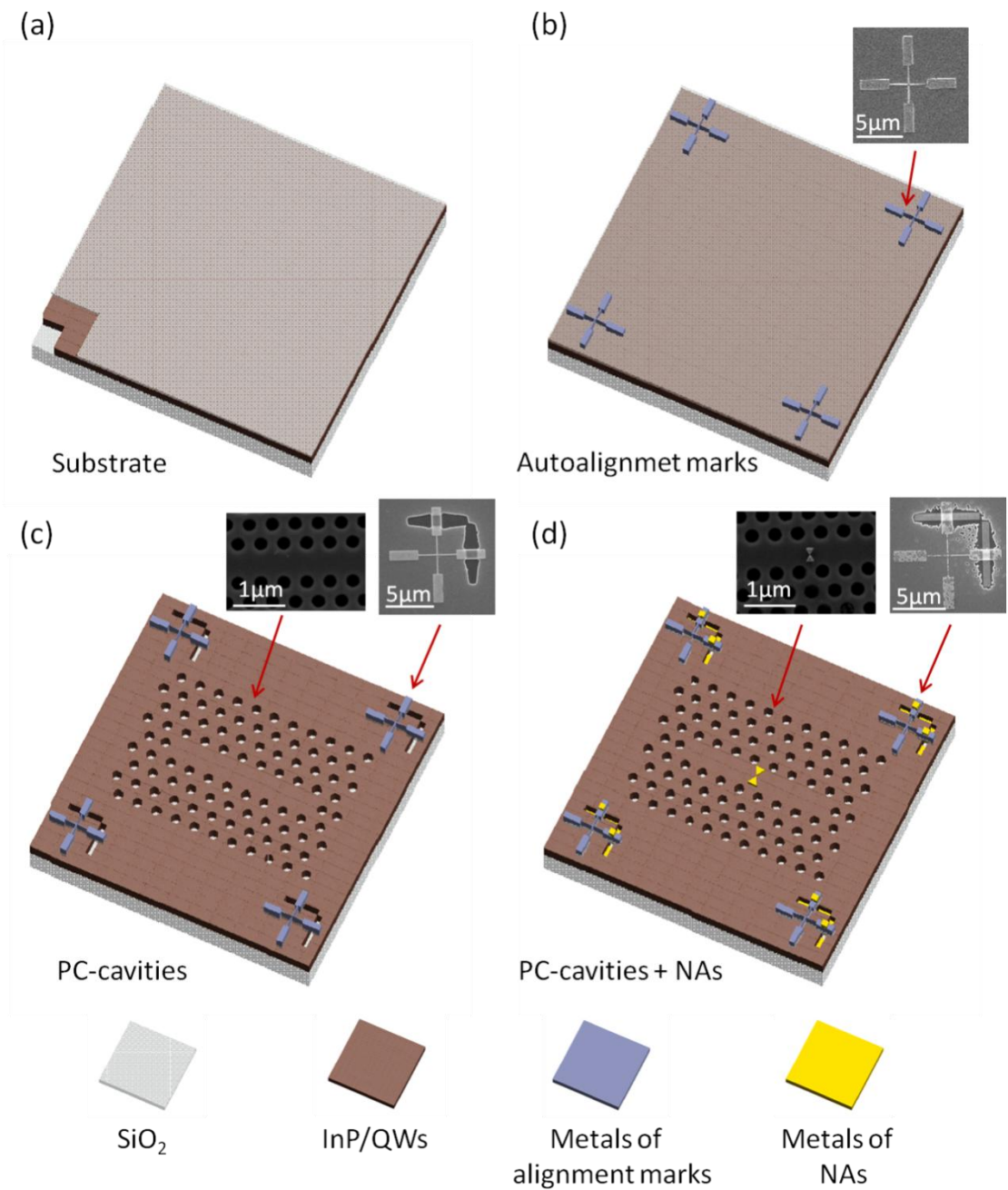

**Figure 3:** Process flow chart of the hybrid nanodevices highlighting the three-step electron-beam lithography process. The relative positions of the gold antenna and the photonic crystal were accurately aligned using gold markers.

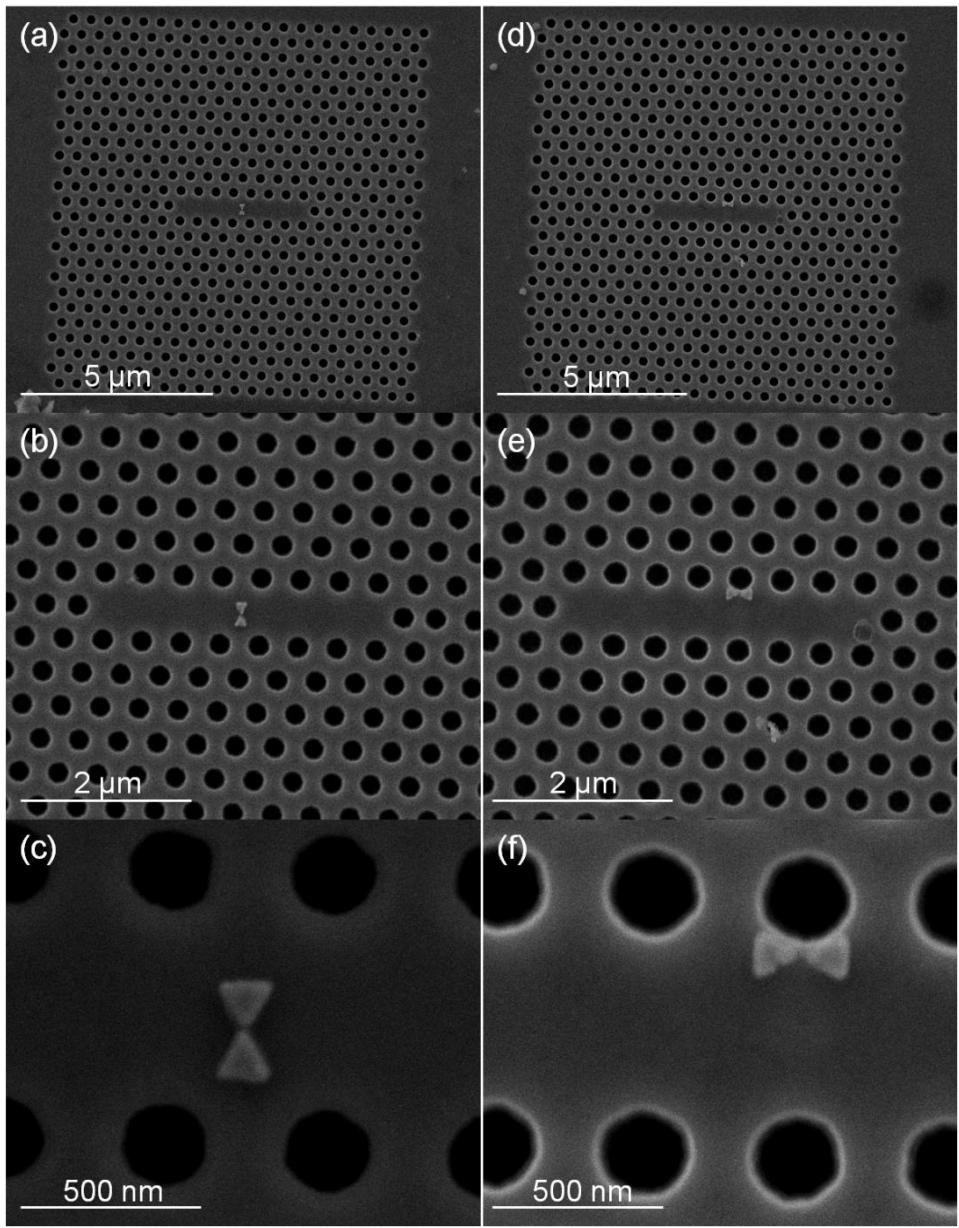

**Figure 4:** SEM image of two hybrid nanostructures: (a), (b) and (c) are the structure with the NA coupling with the Ey field; (d), (e) and (f) are the structure with the NA coupling with the Ex field.

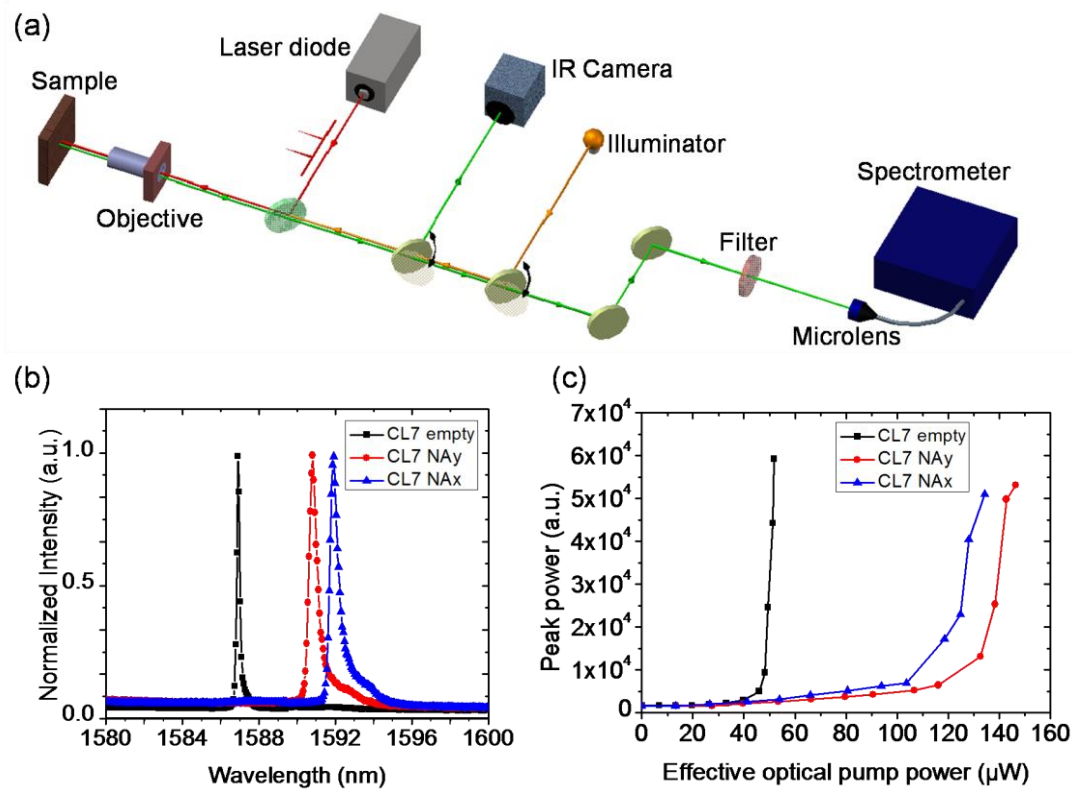

Figure 5: (a) Optical set-up used for spectra measurements; (b) normalized lasing spectra of a group of hybrid structures together with a bare PC-cavity. Spectra have been performed above the threshold for each structure (51.3 μW, 146.5 μW and 134.3 μW for the empty cavity, the CL7-NAy and CL7-NAx structures respectively); (c) variation of the laser peak intensity versus the effective incident pump power of the same group of devices. The lasing thresholds are respectively 45 μW, 130 μW and 100 μW for the empty cavity, the CL7-NAy and CL7-NAx structures) .